\documentclass[
preprint,
amsmath,amssymb,superscriptaddress,aps,%pre+-
prl]{revtex4-1}

\setcitestyle{super,comma,sort&compress}

\usepackage[utf8]{inputenc}
\PassOptionsToPackage{hyphen}{url}
\usepackage{hyperref}
\expandafter\def\expandafter\UrlBreaks\expandafter{\UrlBreaks%  save the current one
  \do\a\do\b\do\c\do\d\do\e\do\f\do\g\do\h\do\i\do\j%
  \do\k\do\l\do\m\do\n\do\o\do\p\do\q\do\r\do\s\do\t%
  \do\u\do\v\do\w\do\x\do\y\do\z\do\A\do\B\do\C\do\D%
  \do\E\do\F\do\G\do\H\do\I\do\J\do\K\do\L\do\M\do\N%
  \do\O\do\P\do\Q\do\R\do\S\do\T\do\U\do\V\do\W\do\X%
  \do\Y\do\Z}

\usepackage{amsmath}
\usepackage{times}
\usepackage{graphicx}
\usepackage{tabularx}
\usepackage{booktabs}

\usepackage{parskip}

\usepackage{color}
\usepackage[normalem]{ulem}
\usepackage{comment}

\begin{document} 

\title{COVID-19 in Africa -- outbreak despite interventions?}
%Containment of COVID-19 in South Africa

% Place the author information here.  Please hand-code the contact
% information and notecalls; do *not* use \footnote commands.  Let the
% author contact information appear immediately below the author names
% as shown.  We would also prefer that you don't change the type-size
% settings shown here.

\author{Malte Schr\"oder}
\affiliation{Chair for Network Dynamics, Cluster of Excellence Physics of Life, Institute for Theoretical Physics and Center for Advancing Electronics Dresden (cfaed), Technical University of Dresden, Helmholtzstr. 18, 01069 Dresden, Germany}

\author{Andreas Bossert}
\affiliation{Center of Methods in Social Sciences, Department of Social Sciences, Georg August University Göttingen, Goßlerstraße 19, 37073 Göttingen, Germany}
\affiliation{Next Generation Mobility (NGM), Department of Dynamics of Complex Fluids, Max-Planck-Institute for Dynamics and Self-Organization, Am Fassberg 17, 37077 G\"ottingen, Germany}

\author{Moritz Kersting}
\affiliation{Next Generation Mobility (NGM), Department of Dynamics of Complex Fluids, Max-Planck-Institute for Dynamics and Self-Organization, Am Fassberg 17, 37077 G\"ottingen, Germany}
\affiliation{Chair of Statistics, Department of Economics, Georg August University G\"ottingen, Humboldtallee 3, 37073 Göttingen, Germany}

\author{Sebastian Aeffner}
\affiliation{Institute for Diagnostic and Interventional Radiology, University Medical Center G\"ottingen, Robert-Koch-Straße 40, 37075 Göttingen, Germany}

\author{Justin Coetzee}
\affiliation{GoMetro, 10 Church Street, Durbanville, Cape Town, South Africa, 7550}

\author{Marc Timme}
\affiliation{Chair for Network Dynamics, Cluster of Excellence Physics of Life, Institute for Theoretical Physics and Center for Advancing Electronics Dresden (cfaed), Technical University of Dresden, Helmholtzstr. 18, 01069 Dresden, Germany}
\affiliation{Institute for the Dynamics of Complex Systems, Faculty of Physics, Georg August University of Göttingen, Friedrich-Hund-Platz 1, 37077 Göttingen, Germany}

\author{Jan Schl{\"u}ter}
\affiliation{Next Generation Mobility (NGM), Department of Dynamics of Complex Fluids, Max-Planck-Institute for Dynamics and Self-Organization, Am Fassberg 17, 37077 G\"ottingen, Germany}
\affiliation{Institute for the Dynamics of Complex Systems, Faculty of Physics, Georg August University of Göttingen, Friedrich-Hund-Platz 1, 37077 Göttingen, Germany}

%%%%%%%%%%%%%%%%% END OF PREAMBLE %%%%%%%%%%%%%%%%

\begin{abstract}
\setlength{\parindent}{0pt}
In Africa, while most countries report some COVID-19 cases, the fraction of reported patients is low, with about  
$20\,000$ 
cases compared to the more than $2.3$ million cases reported globally as of April 18, 2020. 
Few African countries have reported case numbers above one thousand, with South Africa reporting
$3\,034$ 
cases being hit hardest in Sub-Saharan Africa. 
Several African countries, especially South Africa, have already taken strong non-pharmaceutical interventions that include physical distancing, restricted economic, educational and leisure activities and reduced human mobility options.
The required strengths and overall effectiveness of such interventions, however, are debated 
because of simultaneous but opposing interests
 in most African countries:
strongly limited health care capacities 
and testing capabilities largely conflict with pressured national economies and socio-economic hardships on the individual level, limiting compliance to intervention targets.
Here we investigate implications of interventions on the COVID-19 outbreak dynamics, focusing on South Africa before and after the national lockdown enacted on March 27, 2020. Our analysis shows that initial exponential growth of existing case numbers is consistent with doubling times of about $2.5$ days. After lockdown, the growth remains exponential, now with doubling times of 18 days, but still in contrast to subexponential growth reported for Hubei/China after lockdown. 
Moreover, a scenario analysis of a computational data-driven agent based mobility model for the Nelson Mandela Bay Municipality (with $1.14$ million inhabitants) hints that keeping current levels of intervention measures and compliance until the end of April is of insufficient length and still too weak, too unspecific or too inconsistently complied with to not overload local intensive care capacity.
Yet, enduring, slightly stronger, more specific interventions 
combined with sufficient compliance
may constitute a viable option for interventions for regions in South Africa and potentially for large parts of the African continent.
\end{abstract}
%
% currently 292 words *checkmark*
%

%
% also needs: research in context
%
%Research in context
%Evidence before this study This section should include a description of all the evidence that the authors considered before undertaking this study. Authors should briefly state: the sources (databases, journal or book reference lists, etc) searched; the criteria used to include or exclude studies (including the exact start and end dates of the search), which should not be limited to English language publications; the search terms used; the quality (risk of bias) of that evidence; and the pooled estimate derived from meta-analysis of the evidence, if appropriate.
%Added value of this study Authors should describe here how their findings add value to the existing evidence.
%Implications of all the available evidence Authors should state the implications for practice or policy and future research of their study combined with existing evidence.
%Research in context panels should not contain references; key studies mentioned here should be referenced in the main text.
%

\maketitle 

\newpage
%\section*{Research in Context}
% SEBASTIAN AEFFNER: 
%- please check "Contributors" far below, it needs to be final before we collect author signature, please check your email around lunch time
%wenn Du Lust hast, kannst Du an dieser Research in Context section schreiben, sowas habe ich noch nie gemacht (Marc). Three parts: evidence before study, added value of this study, implications of all available evidence,
%Research in context panels should not contain references; key studies mentioned here should be referenced in the main text.

%%official description
%%Putting research into context•    All    research    papers    (including    systematic    reviews/meta-analyses)  submitted  to  any  journal  in  The  Lancet  family  must  include a panel putting their research into context with previous work in the format outlined below (see Lancet 2014; 384: 2176–77,  for  the  original  rationale).  This  panel  should  not  contain  references.    Editors    will    use    this    information    at    the    first    assessment stage and peer reviewers will be specifically asked to check the content and accuracy•  The  Discussion  section  should  contain  a  full  description  and  discussion  of  the  context.  Authors  are  also  invited  to  either  report  their  own,  up-to-date  systematic  review  or  cite  a  recent  systematic review of other trials, putting their trial into context of the review

\section*{Executive Summary}

\textbf{Evidence before this study.} Several studies have analyzed the ongoing outbreak of the Corona Virus Disease 2019 (COVID-19) in China and several European countries. However, conditions in African countries are vastly different and often fragile, with conflicting limitations of both the health care system and socio-economic conditions, posing difficult challenges for decisions about enacting and lifting interventions. These countries are currently in the early stages of the outbreak and have been reporting a small but rapidly increasing number of patients diagnosed with COVID-19. Several countries have taken different intervention measures to counter a large-scale COVID-19 outbreak. In particular, in South Africa, with the largest number of cases in Sub-Saharan Africa, case numbers are known to less rapidly increase after national lockdown on March 27, 2020.

\textbf{Added value of this study.}
This study reports a quantitative analysis of the case number dynamics reported by the World Health Organization and Johns Hopkins University until including April 18, 2020, both for Africa overall and South Africa specifically, before and after national lockdown. It also reports and analyzes results of an agent-based mobility simulation for the Nelson Mandela Bay Municipality, South Africa ($1.14$ million inhabitants). This case study relies on detailed large-scale mobility survey data of about 10\% of the population and on estimates of the fractions by which interventions decrease specific activities. The simulational data on outbreak dynamics thus 
%do not deliver exact quantitative predictions, yet 
provide qualitative order of magnitude estimates of trends consistent with past data. Combined, both analyses may help to better understand the implications of interventions on and estimate the dynamics of the number of (critically) infected patients. 

\textbf{Implications of all the available evidence.}
The results suggest that current interventions are not yet sufficient to contain a larger-scale outbreak. Interventions slightly stronger than those implemented today or a higher degree of compliance to the enacted lockdown, in combination with longer-lasting measures than currently announced for South Africa may help bound the case numbers such that the number of critical patients remains at or below (and does not massively overburden) the local capacity of intensive care units. Strategies for strengthening or lifting  interventions should be advised by advanced data analytics and predictive modeling estimates, 
%combining advanced data analytics with agent-based modeling  may offer a useful tools to advise strategies for strengthening or lifting non-pharmaceutical interventions
for instance for evaluating necessary time intervals and required levels of interventions. Overall, the study points to a potentially viable chance for effective non-pharmaceutical countermeasures against COVID-19 epidemics in South Africa, with suggestions for Health Policy for large parts of the African continent and, generally, disadvantaged countries and regions.

\newpage
\section*{Introduction}
The severe acute respiratory syndrom coronavirus 2 (SARS-CoV-2) has reached more than 200 countries and territories across all continents \cite{WHO56, JohnsHopkins2020}. By death toll, the resulting Corona Virus Disease 2019 (COVID-19) outbreak will likely soon become the largest pandemic of the 21st century so far \cite{callaway2020coronavirus}. 
There is currently no specific medical intervention known against SARS-CoV-2 and preventive vaccination options are not yet available. %may only be available in 2021 or later \green{reference(s) needed}.
%The resulting vast number, broad geographical distribution and strengths of socio-economic interventions may be unprecedented in modern human history. 
The resulting vast number, broad geographical distribution, and intensity of globally enacted socio-economic interventions is unprecedented in modern human history. 
 
Mainland China was the first region hit by the outbreak in January 2020 and had taken rapid and severe interventions including an almost complete lockdown for eleven weeks. It thereby succeeded to suppress the outbreak dynamics to subexponential growth patterns \cite{maier2020effective} and in April 2020 is reporting a total of $83\,500$ cases and at most 130 new cases daily for now more than four weeks.\cite{JohnsHopkins2020} As of April 18th, several countries in Europe are reporting more than $100\,000$ cases each and the United States alone reports above $700\,000$ cases. %while all other American countries combined as well as the entire Western Pacific Area including Australia report between 130,000 and 200,000 total \cite{WHO_covidashboard} \green{I would suggets to remove the part after "while": Situation in Western Pacific/Australia does not seem to be relevant for the purpose of the paper}. 
At the same time, Africa as a continent with a population of $1.3$ billion people as of 2018\,\cite{un2019population} %\green{more recent value?} 
has reported only about %11843 
%12000
$21\,000$ 
cases and about %800 
$1\,000$ %(approximately 6.7\%)
(approximately $4.8$\%) deaths. \cite{WHO_covidashboard, africaCDC_outbreakbrief13, JohnsHopkins2020} Of those, the largest number of COVID-19 patients is reported in South Africa with %2608 
$3\,034$ 
cases and %48 (about 2\%) 
52 (about $1.7$\%) 
deaths as of April 18, 2020.\cite{sacoronavirus, JohnsHopkins2020} 
Across all these countries, the total number of cases is rapidly increasing.  

Across the African continent, national economic constraints, individual poverty, low health literacy rates, weaker health care systems and cultural practices lead to reduced option spaces on personal and governmental levels and may all contribute to more severe consequences of the COVID-19 outbreak and negatively influence containment as well as recording, testing and medical treatment \cite{velavan2020covid}. Similar conditions will hold for most countries of the Global South, calling for particular attention on African countries \cite{nkengasong2020looming}.

%XX WHERE TO PUT THIS? \blue{Here?}
%South Africa is one of the African countries best connected to Mainland China \cite{Gilbert2020}, which put it at a high risk for an early and extensive import of several diseases that caused epidemics. These connections notwithstanding, most initial COVID-19 cases reported in early March, 2020, refer to people returning from Italy and other parts of Europe. \blue{WHY THE CHINA/ITALY PART?} 

In general, health care systems in African countries feature only a small number of available intensive care units (ICUs) compared to most countries of the Global North.\cite{WHO_africaReadiness, murthy2015intensive} At the same time, African countries are under particular pressure due to 
economic constraints on both national and personal levels. Besides strong repercussions on national economic productivity expected for any large-scale lockdown, a large fraction of the population is unable to fully comply with severe lockdown measures due to their personal financial situation.
An African task force for coronavirus preparedness and response (AFTCOR) has been established to manage these combined and conflicting constraints both for the current COVID-19 outbreak and for future preparedness \cite{hopman2020managing}. Their work focuses on enabling medical diagnosis and screening options, clinical treatment of COVID-19 patients, infection prevention and control in health care facilities, supply chain management, and the communication of risks to experts and the public. Qualitative and quantitative data analytics and estimates of the outbreak dynamics and evaluating containment options essentially underlie but are not in the focus of their work.

South Africa offers a comparatively high capacity of intensive care units (ICUs) to respond to outbreaks, with estimates ranging from maximally $7\,195$ ICU beds theoretically in existence to $2\,926$ practically available nationwide across both public and private sectors \cite{vandenheever2020}. The order of magnitude of these numbers is consistent with earlier reports.\cite{bhagwanjee2007ICUavailabilitySouthAfrica} However, the factually available ICU beds have likely declined during the past decade necessitating rationing and triage (prioritisation) decisions that have been frequently necessary in South Africa even in times before COVID-19, particularly in the publicly funded health sector.\cite{vandenheever2020, joynt2019critical} Moreover ICU capacity in the private sector is not readily and generally accessible.

\section*{Influence of lockdown on past case numbers}
On March 5, 2020, the first COVID-19 patient has been confirmed in South Africa and after starting with specific smaller measures from March 15 onwards, the South African government enacted a national lockdown effective March 27, 2020. This lockdown includes measures such as the complete closure of \textit{childcare}, institutions of primary and higher \textit{education} as well as all public \textit{leisure} activities, severe \textit{physical distancing} rules, an estimated 70\% reduction of \textit{shopping}, 85\% of on-site \textit{work} force and a 90\% reduction in \textit{other} activities.
An initial formal reduction of shared publicly available mobility services by about 75\% was, after protests, revised to about 30\% reduction\cite{CNBCAfrica2020MinibusTaxiLockdownRevised} (estimates by GoMetro, South Africa). These shared mobility services provide a large fraction of transportation and constitute one of the special conditions in South Africa and many other African countries.\cite{govender2016MinibusSerivcesSA} For instance in South Africa, instead of formal public transit, transportation is dominated by private, semi-regulated minibus taxis with typically 15 seats.\cite{govender2016MinibusSerivcesSA} Due to their mass usage, usually high occupancy and the close contact between passengers in the vehicles, these mobility services may contribute substantially to the spread of COVID-19.

%Fig 1 main finding here:
Fitting the number of total reported cases in South Africa before and after the national lockdown (Figure 1) indicates that the lockdown drastically reduces the relative increase in case numbers, as quantified by the growth exponent, decreasing from $r=0.32$ per day in the beginning of the outbreak to about $r=0.27$ per day just before the lockdown and down to $r=0.038$ per day after the lockdown, reflecting an increase of the doubling time from about $2.5$ to about $18$ days (Fig.~1, panels A and B). The immediate switch to slower growth at the date of the official lockdown may be originating from several factors that remain unknown. For instance, the number of patients tested per day has substantially increased initially \cite{SACOronaStats_Wiki, sacoronavirus} and tests may have been delayed at the very onset. In any given region, the first person infected is likely detected only after exhibiting symptoms while later cases may be identified by preemptive contact tracing and thereby identified as they appear, ideally before showing symptoms. Other contributing factors may include stochastic small number fluctuations occurring at the onset of any epidemic outbreak,  already existing awareness of the COVID-19 outbreak and many individuals taking partial countermeasures before the official national lockdown.

As the number of cases in South Africa makes up a substantial share of all reported cases throughout Africa, the effect also becomes visible in the data for the entire continent (Fig.~1A,B). For Africa as a whole, growth exponents dropped from about $r=0.22$ to $r=0.086$. The data for Africa suggest a further decrease of the exponent, ongoing after the South African lockdown. This downtrending may be associated with measures taken up at different points in time in the most strongly affected countries of Northern Africa, and the vastly heterogeneous case numbers, test coverage and reporting of cases across African countries, all entangling with the reduced number, but still large share of South African COVID-19 patients.

While the growth exponents have been substantially reduced, between a factor of $7.1$ (South Africa) and a factor of $2.6$ (all of Africa), the growth remains exponential, even three weeks into the lockdown. This is in stark contrast to the outbreak dynamics in Mainland China, where the strict containment measures of the Hubei region has led to subexponential growth \cite{martin2020effectiveness} followed by a massive decrease of new case numbers within weeks after lockdown \cite{JohnsHopkins2020}. The unbroken exponential growth trend in South Africa is also indicated by the number of newly infected people per week steeply increasing when displayed as a function of the total number of infected (Figure 1C), instead of curving down.

\section*{Modeling future scenarios}

% results of Fig 2, later also of Fig 3.
The current national lockdown has been extended from an original three weeks (until April 17, 2020) with relaxations now suggested for the beginning of May, 2020. We thus ran scenario simulations to estimate future case numbers and probe responses to different intervention strengths and durations. We employed a computational data-driven, agent based transport model for the Nelson Mandela Bay Municipality (NMBM, Eastern Cape, South Africa, $1.14$ million inhabitants) \cite{bossert2020limited} with lockdown fractions of work, leisure, and shopping activities and complete lockdown of childcare and educational institutions, in line with measures currently implemented in South Africa. To reflect potential non-compliance with enacted lockdown measures, the simulations took only a 85\% reduction of \textit{other} activities; for minibus taxi services we took a 50\% effective reduction of passengers, to reflect the tradeoff between non-compliance and the reduction in demand due to less people required or wishing to travel caused by the other lockdown measures and the outbreak (estimates by GoMetro, South Africa).

%to reflect the reduction in demand due to less people required or wishing to travel caused by the other lockdown measures (estimates by Justin Coetzee, GoMetro, South Africa).

Our simulations are consistent with the growth exponents of the total number of infected individuals both before and after national lockdown, indicating growth factors of about $r=0.32 \pm 0.01$ %\red{where does the 0.03 come from? My data says 0.323 +/- 0.003 for the fit to the average (plotted) and \textbf{0.32 +\- 0.01} for the average over all fits. We should give the second number as this is also the number we report for the default lockdown} 
and about $r=0.04 \pm 0.02$, %\red{correct}, 
respectively (Figure 2A,B). The exponents cannot be specified more exactly due to the unpredictable stochastic factors in the transmission process creating substantial variations in particular at low case numbers, sampled over in simulations with ten random realizations each. Importantly, there are simulated case dynamics that display an early (within April, 2020)  saturation of the total number of cases at $10\,000$ or below. However, the ensemble of simulations of the lockdown scenario suggests an ongoing outbreak either entirely without saturation or with early but non-persistent saturation and renewed increase in May. Figure 2C displays the same data of the dynamics in a state space characterizing the epidemics without referring to absolute time (as in Fig. 1C), thereby enabling to compare system-wide potential pathways. The results illustrate that current lockdown measures substantially slow the spread of the outbreak in all realizations, but only 1 out of 10 realizations manages to completely stop the outbreak while hundreds of thousands of people become infected in the other realizations in the Nelson Mandela Bay Municipality alone.

% results of Fig 3
To evaluate the expected outbreak dynamics and the maximal number of critical patients requiring intensive care, we studied four different scenarios by agent-based simulations, again ten realizations per scenario (Figure 3). Entirely lifting the currently enacted lockdown on May 1 would cause an immediate rise of infected patient numbers and a delayed rapid rise of critical patient numbers drastically beyond the ICU capacity available in NMBM (estimated to be $50$ based on downscaling (proportional, by population size) the $267$ ICU beds expected to be available in the entire Eastern Cape Province \cite{SAMJ6415}).
%, wiki_eastern_cap, wiki_mnmb}). %\green{(difficult to understand)}. 
%\blue{(out of the $267$ ICU beds expected in the entire Eastern Cape Province \cite{SAMJ6415}, about $50$ would be available for NMBM proprtional to its population share) [better to understand?]}
Whereas the exact numbers will depend on details of the simulation, further simulations (not shown) indicate a manifold overload of ICU capacity also after varying mobility parameters. Lifting lockdown by 25\% two weeks later, i.e. on May 15, still would cause massive rise in case numbers and  ICU overload in early June. Maintaining current lockdown conditions strongly slows the outbreak, yet our simulations suggest that such interventions together with current compliance are marginally insufficient to contain the epidemic long term and keep the number of critical patients below ICU capacity (Fig.~3B and C), as suggested already by our data analysis of past case numbers (Figure 1). Finally a fourth scenario of slightly strengthening current interventions, either by slightly stricter, possibly even more specific lockdown regulations, by increasing compliance, or a combination both (90\% reduction of shopping and other, 95\% reduction of work activities and complete restriction of all other activities including public mobility in the simulations), may keep the number of critical COVID-19 patients at or below the ICU capacity and may largely contain the epidemic by end of June 2020.

%The capacity for Number of ICU capacity
%compare also  \cite{bossert2020limited}.
%reduced number of activities, in particular a lower number of people commuting to or for work ,substantially reduces the factual usage of shared public mobility services, 

\section{Discussion}
The analysis of reported past case data is robust and suggests that the outbreak currently still grows too quickly to contain the number of critical COVID-19 patients significantly below available ICU capacities nation-wide. The main potential causes of errors in the analysis of past data may be biased or undersampled testing and reporting of case numbers. 

Predicting future case numbers and the number of critical patients under different scenario conditions is much more difficult. The most difficult challenge is the bridging of scales between known or estimated country-wide overall conditions and specific urban level scenarios (at $1.14$ million people) that are again subsampled at about 10\% of the population, \textit{not} primarily due to simulational constraints but due to the availability of socio-economic and travel data for about $100\,000$ people only.\cite{bossert2020limited} Combined with the COVID-19 outbreak being at an early stage, the number of infected patients is of an order of magnitude between $10^1$ and $10^3$ in NMBM, thereby causing strong stochastic number fluctuations that make individual predictions unreliable. We attempted to compensate for such fluctuations partially by running ensemble simulations for 10 random realizations, with a random subsample of initial patients infected (and thus varying their location, household size, employment status etc.). As the results are based on limited ensemble simulations, they likely underestimate the probability of extreme outcomes such as strong increase or random decay of the outbreak.

%\blue{delete here and move to conclusion? otherwise doubled}
%Across the African continent, national economic constraints, individual poverty, low health literacy rates and cultural practices lead to reduced option spaces on personal and governmental levels and may all \out{negatively influence} \blue{contribute to more severe consequences of the} COVID-19 outbreak \blue{and negatively influence} recording, testing and medical treatment. Similar conditions will hold for most countries of the global South.

\section{Conclusion}
The results reported above suggest that current lockdown levels may be just marginally insufficient to prevent a massive COVID-19 outbreak in South Africa.
As the increase in case numbers is still exponential and not subexponential as reported for Mainland China \cite{martin2020effectiveness}, South Africa may be still in the unfortunate situation to become for the African continent what Italy has been for Europe\cite{REMUZZI2020}, with potentially devastating consequences.

A rapid large-scale infection within weeks to a few months, the likely outcome if the national lockdown was lifted or relaxed early May\cite{sacoronavirus}, implies a manifold overload of ICU capacity.
Interventions slightly stronger than those implemented today, or even a higher degree of compliance to the enacted lockdown alone may constitute a viable chance for effective countermeasures for regions in South Africa and potentially for large parts of the African continent.

However, a number of boundary conditions beyond those known for past major hubs of the COVID-19 pandemic in countries like Mainland China, the United States or Italy \cite{REMUZZI2020} need to be taken into account simultaneously. Most African countries find themselves under much stronger socio-economic and health care system constraints than countries of the Global North. 

For instance, a large fraction of the work force both is at lower-income levels and simultaneously has no fall-back option to remote work. As many of such work activities are not tagged "essential" in the sense of the lockdown, people often have zero income or immediately fall into extreme poverty. Moreover, even where remote work is possible, it comes with additional challenges \cite{Watson2020RemoteWork}. Still, South Africa is potentially in a better position than many other African countries, so the conclusions (for South Africa specifically) might be conservative in this sense.

The South African health situation includes a high risk of COVID-19 coinfections for patients with, e.g., HIV/AIDS or forms of tuberculosis (TBC) and implies additional challenges.
%\green{which may lead to poorer outcomes 
%[cite e.g. "World Health Organization (WHO) Information Note Tuberculosis and COVID-19"} \red{and the special risk for other patients, e.g. those under oncological treatment} 
%\green{[I'd rather remove the oncology part (too unspecific) and focus on HIV and TBC -- These seem to be particularly high in SA (see next sentence), while cancer rates in SA are most likely comparable to other countries]}. 
According to the WHO 2019, South Africa ranks 4th globally in the number of TBC infections \textit{per capita} and 3rd for those coinfected with TBC and HIV. Moreover, the South African population infected with TBC alone is about $320\,000$ ($0.5\%$, about 20 times higher rate than in Europe) and a total of $7.7$ million people ($13\%$) are infected with HIV as of 2018.\cite{world2019global, UNAIDS2018HIV} 

Therefore, regulatory decision against COVID-19 cannot only keep in mind short-term economic constraints.\cite{CNBCAfrica2020MinibusTaxiLockdownRevised} A large-scale outbreak and massive ICU overload may have drastic consequences for the country as a whole, including societal and economic but also  psychological, and ethical issues (compare \cite{Ramathuba2020EthicalConflictsSouthAfrica}). Thinking of economic constraints should also imply thinking of long-term implications, for both economy and society. 
This perspective underlines again the coaction advocated by the United Nation's Sustainable Development Goals (SDGs), in particular Good Health and Wellbeing (SDG 3), Sustainable Cities and Transportation (SDG 11), and Reduced Inequality both within and among countries (SGD 10) in the context of COVID-19. 

An integrated perspective on such goals may help paving the way to a fair and sustained solution of the COVID-19 crisis and future pandemics across African countries as well as for individuals, groups and regions in a position much more fragile than common for countries of the Global North, as also underlined by the proposed CoHERE programme.\cite{Horton2020CoHEREpostPandemicHealthStragegy}

%The ratios of public ICU beds to population varies drastically throughout South Africa, from the best at 1:14 000 to 1:20 000 in the Western Cape, to the worst at 1:82 000 to 1:150 000 in Limpopo \cite{bhagwanjee2007ICUavailabilitySouthAfrica, joynt2019critical}. Moreover,  

%[19,20] These ratios probably under-represent the disparity as they did not account for the likely use of private services by the more affluent urban populations of the Western Cape and Gauteng. Despite the limited access of the majority of the population to private sector ICU care

%work force remote work already Italy problem 50\% of cannot work remotely from
 %not least regarding medical and care personell, not on

%references from the supplement (we may not need to include them)\
%IREELEVANT NOW, SINCE WE HAVE NO SUPPLEMENT
%
%\nocite{Horni.2016,JOUBERT20181012, neumann2015, bischoff2017, neumann2015, gtfs2011, algoabus, mueller2020, Kermack.1927, Keeling.2008, Anderson.2010, Roche.2011, mueller2020}

\bibliographystyle{short_author_unsrt}
\bibliography{references}

\subsection*{Funding}
MS and MT acknowledge support by the German National Science Foundation (Deutsche Forschungsgemeinschaft, DFG) and the Saxonian State Ministry for Higher Education, Research and the Arts under Germany's Excellence Strategy – EXC-2068 – 390729961 – Cluster of Excellence \textit{Physics of Life} (PoL) and the Center for Advancing Electronics Dresden (cfaed). 

\subsection*{Role of Funding Sources}
The funders had no role in study design, data collection, data analysis, data interpretation, writing of the article, or the decision to submit for publication. All authors had full access to all the data reported in the study and were responsible for the decision to submit the Article for publication.

\subsection*{Contributors}  
MS, MT and JS conceived and designed research. MS and MT worked out and evaluated the analysis driven by past case data. AB and MK designed, set up and adapted the simulation software and ran simulations, supervised by JS. AB, MK and JS evaluated the simulation data, advised by MS and MT. MS and MT provided theoretical background and advised on general data analysis and data presentation. SA advised on health care data and provided medical background. MS and MT wrote the basic version of the manuscript. JC provided local data and advised on conditions for mobility simulations. All authors interpreted the results and contributed to revising and editing the manuscript. 

\subsection*{Declaration of interest} We declare no competing interests. 
 
\clearpage

\normalsize
\setlength{\baselineskip}{6mm}

{
\centering
\includegraphics[width=95mm]{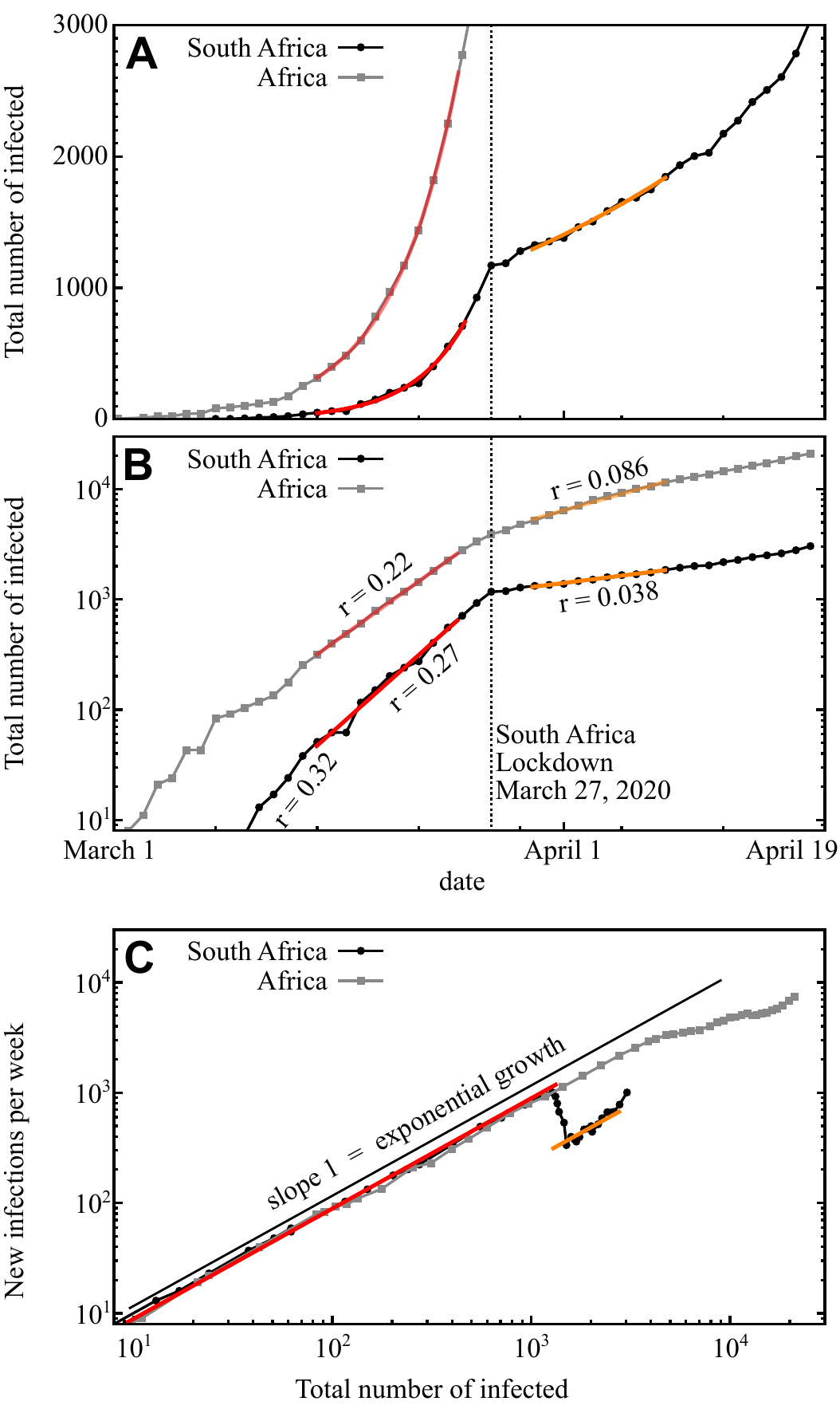}\\
}
\noindent {\bf Fig. 1. COVID-19 in Africa and South Africa}. (A) and (B) The number of confirmed COVID-19 patients in Africa (gray squares) and specifically South Africa (black disks) from March 1, 2020 until April 16, 2020, on (A) linear  and (B) logarithmic scales. Best exponential fits (colored lines) yield growth rates $r$ where the total number of (reported) infected patients $N \propto \exp(r \, t)$ where $t$ measures time in days. (C) State space representing the number of newly reported patients as a function of the total of reported people infected (including the recovered), eliminating absolute time. Straight solid line of slope 1 indicates pure exponential growth
%\red{that is rapid compared to the duration of the infection on any given patient. [What is this supposed to mean?]} 
The impact of the lockdown executed on March 27 is clearly visible (vertical lines in panels A and B).

%
% fits for 
% 15.03 - 25.03 (before lockdown) [day 75 to 85]
% 30.03 - 0.9.04 (after lockdown) [day 90 to 100] (also for Africa, though there the slope changes more, likely due to other countries taking measures later or not keeping up with the reporting)
%

\begin{comment}

\newpage

\begin{minipage}[c]{107mm}
\centering
\includegraphics{FIG1_africa_data_v3.pdf}\\
\end{minipage}
\begin{minipage}[c]{10mm}
\null
\end{minipage}
\begin{minipage}[c]{55mm}
\noindent {\bf Fig. 1. COVID-19 in Africa and South Africa}. (A) and (B) The number of confirmed COVID-19 patients in Africa (gray squares) and specifically South Africa (black disks) from March 1, 2020 until April 16, 2020, on (A) linear  and (B) logarithmic scales. Best exponential fits (dashed lines) yield growth rates $r$ where the total number of (reported) infected patients $N \propto \exp(r \, t)$ where $t$ measures time in days. (C) State space representing the number of newly reported patients as a function of the total of reported people infected (including the recovered), eliminating absolute time. Straight solid line of slope 1 indicates pure exponential growth
%\red{that is rapid compared to the duration of the infection on any given patient. [What is this supposed to mean?]} 
The impact of the lockdown executed on March 27 is clearly visible (vertical lines in panels A and B).
\end{minipage}

\end{comment}

\newpage

{
\centering
\includegraphics[width=95mm]{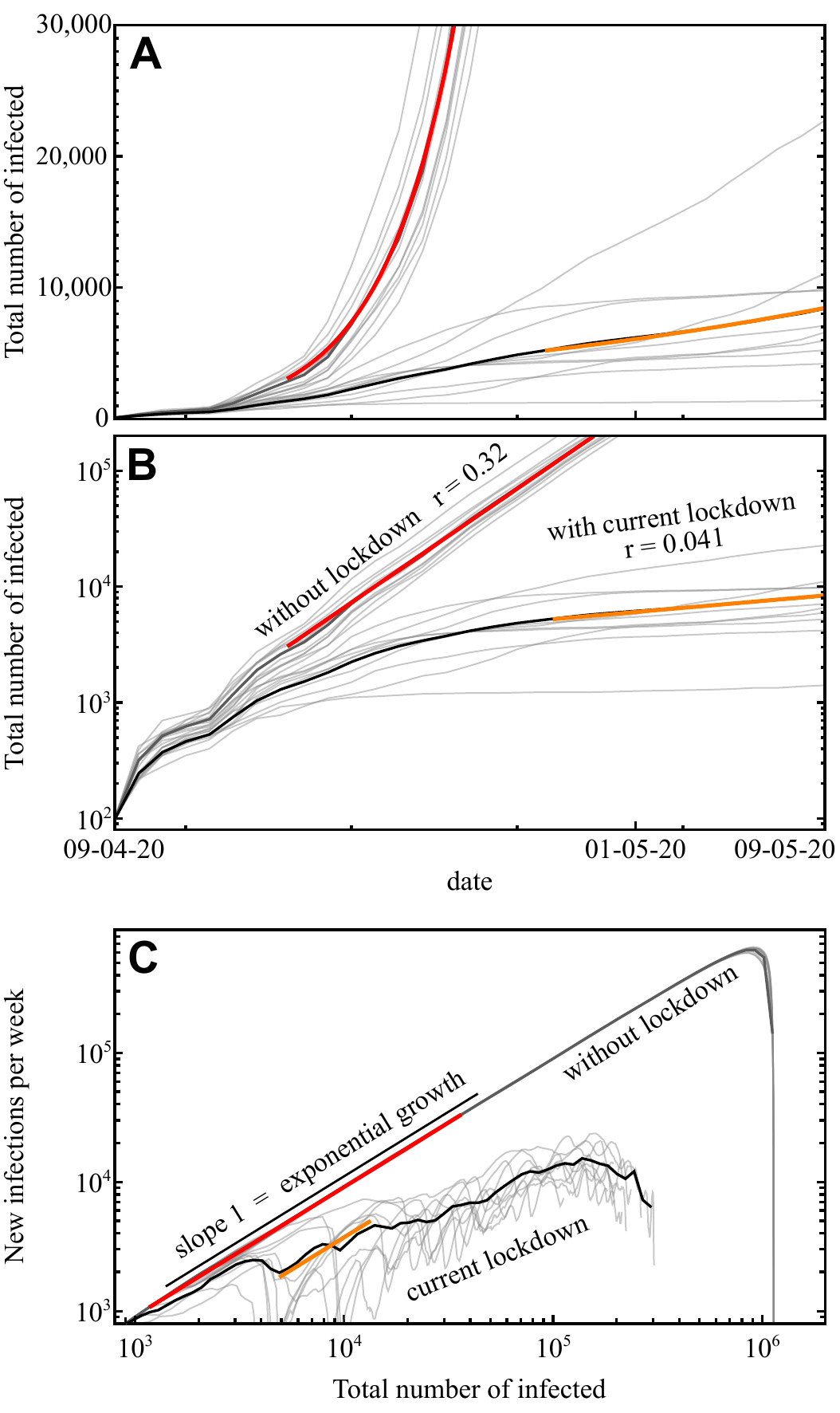}\\
}

\noindent {\bf Fig. 2. Estimated COVID-19 cases for the Nelson Mandela Bay Municipality, South Africa.} 
(A, B) Simulation of the outbreak without interventions (red fit) and with current interventions (orange fit) on (A) linear and (B) logarithmic scales. Thin grey lines represent individual simulations, the solid black lines represent their averages. Growth rates are consistent with the observations for South Africa in the beginning of the outbreak (without lockdown) and after the lockdown (compare Figure 1). 
(C) State space representing the number of newly reported patients as a function of the total of reported people infected (including the recovered), eliminating absolute time. While the lockdown measures slow the spread of the outbreak, the growth remains exponential for some time (compare also Figure 1C).
%
%starting from 100 infected (and infectious) individuals on 09-04-20 with no countermeasures (reproducing slope observed before lockdown) and with currently active countermeasures (reproducing slope observed after lockdown, at least for a part of the trajectory before saturation sets in).
%
%range of growth rates:
%without lockdown 0.3113 to 0.3378 (stddev about 0.01)
%with lockdown 0.01043 to 0.06923 (stddev about 0.02)

\newpage
{
\centering
\includegraphics[width=95mm]{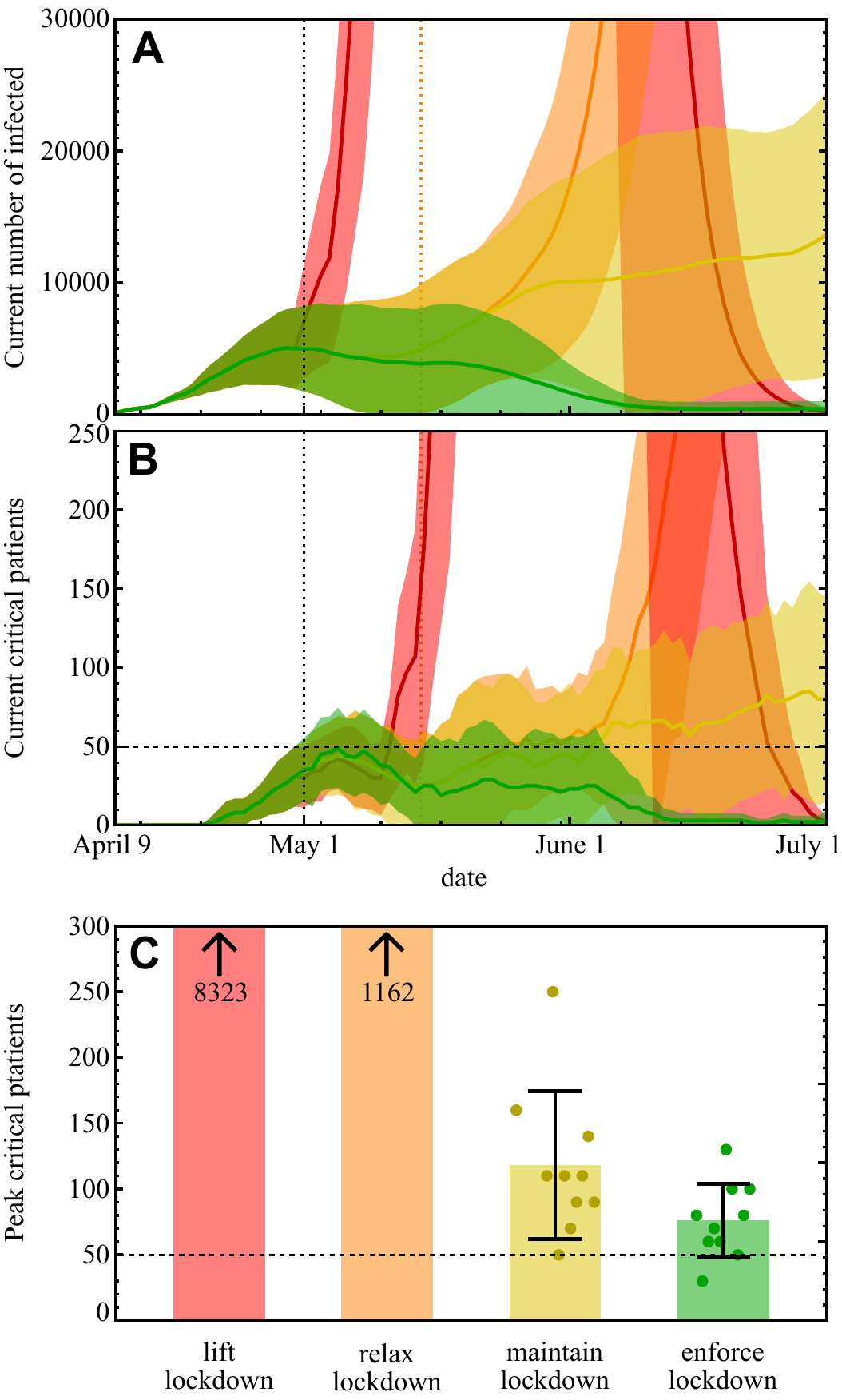}\\
}
\noindent {\bf Fig. 3. 
%Influence of keeping, enforcing and lifting interventions.
Influence of intervention policies
}
(A)  Number of active infections over time. Solid lines indicate averages across realizations, shaded areas indicate standard deviation.
Color encodes the four scenarios: lockdown lifted May 1st (red), lockdown relaxed by 25\% on May 15 (orange), maintain current lockdown until June 30 (yellow), and enforcing lockdown or increasing compliance from May 1, 2020 (green).
(B) Number of critical patients and estimated capacity available in NMBM (horizontal dotted line). Data encoding as in (A). The dashed vertical line illustrates the available ICU capacity.
(C) Maximum number of patients requiring intensive care during the outbreak until end of June 2020, across scenarios (color code as before). Bars indicate averages across realizations and standard deviation, small disks individual realizations. 
Note that these numbers may increase after June 2020, for example when maintaining the lockdown (compare upwards trend in panel B). 
All data based on ten realizations of agent-based simulations for each of four scenarios for NMBM, South Africa. 

%scenarios: lift lockdown in May, maintain lockdown in May or strengthen lokcdown in May (using Justin's adjusted realistically possible lockdown scenario)
%all data up to end of june
%All data based on 10 realizations of agent-based simulations for each scenario for NMBM, South Africa.
\begin{comment}
additional info:
all ten simulations of the enforced lockdown end before the end of july (not shown/visible in the figures)

results for the 3 lockdown scenarios showing the peak critical patients until end of june:

default lockdown (keep the adjusted values after Justin's update)

70, 110, 110, 110, 50, 90, 90, 160, 250, 140
Mean:    118
Median:    110
Mode:    110
Standard Deviation:    56.1348

end lockdown  (back to baseline)

7870, 7510, 8580, 8460, 8380, 8990, 8080, 8290, 8760, 8310
Mean:    8323
Median:    8345
Mode:    7870, 7510, 8580, 8460, 8380, 8990, 8080, 8290, 8760, 8310
Standard Deviation:    428.228

enforce lockdown (Justin's new realistic lockdown)

60, 50, 80, 80, 30, 100, 60, 70, 130, 100
Mean:    76
Median:    75
Mode:    60, 80, 100
Standard Deviation:    28.7518
\end{comment}

\end{document}